\documentclass[aps,prl,twocolumn,showpacs,a4paper,superscriptaddress,groupedaddress]{revtex4}
\usepackage{graphicx}
\usepackage[toc,page]{appendix}
\usepackage{multirow}
\usepackage{dcolumn}
\usepackage{bm}
\usepackage{amssymb}
\usepackage{amsmath}
\usepackage{nonfloat}
\usepackage{natbib}

\newcommand\scalemath[2]{\scalebox{#1}{\mbox{\ensuremath{\displaystyle #2}}}}

\begin{document}

\title{Polarisation control of optically pumped terahertz lasers}
\author{G. Slavcheva}
\affiliation{Blackett Laboratory, Imperial College London,\\
Prince Consort Road, London SW7 2AZ, United Kingdom}
\email{g.slavcheva@imperial.ac.uk}
\affiliation{Mediterranean Institute of Fundamental Physics, Via Appia Nuova 31, 00040
Rome, Italy}
\author{A. V. Kavokin}
\affiliation{Spin Optics Laboratory, St. Petersburg State University, 1, Ulianovskaya,
198504, Russia and School of Physics and Astronomy, University of
Southampton, Highfield, Southampton SO17 1BJ, United Kingdom}

\begin{abstract}
Optical pumping of excited exciton states in semiconductor quantum wells is
a tool for realisation of ultra-compact terahertz (THz) lasers based on
stimulated optical transition between excited ($2p$) and ground ($1s$)
exciton state. We show that the probability of two-photon absorption by a $2p
$-exciton is strongly dependent on the polarisation of both photons. Variation of the threshold power for THz lasing by a factor of
5 is predicted by switching from linear to circular pumping. We calculate the polarisation
dependence of the THz emission and identify photon polarisation
configurations for achieving maximum THz photon generation quantum
efficiency.
\end{abstract}

\maketitle





\textit{Introduction}.-Excitons in nanoscale semiconductor materials exhibit low-energy excitations
in the range of the exciton binding energy, analogous to inter-level
excitations in atoms, yielding infrared and terahertz (THz) transitions.
Thus excited exciton ladder states represent a natural system for generating
THz radiation and coherence. The demand for development of new compact and
efficient coherent terahertz radiation sources is currently rapidly
increasing, due to ever growing range of very diverse technological
applications in the relatively little-explored THz spectrum of radiation
\cite{Tonouchi}. Towards this goal recently a new scheme of a microcavity
based polariton triggered THz laser (THz vertical cavity surface emitting laser
(VCSEL)) has been proposed by one of the authors \cite{Kavokin}, whereby the
$2p$ dark quantum well (QW) exciton state is pumped by two-photon absorption
using a $cw$ laser beam.

In this Letter we theoretically demonstrate polarisation control of THz
emission and of the quantum efficiency for THz photon generation. We
consider a THz VCSEL proposed in Ref.( \cite{Kavokin}), where the pump beam
is split in two. Each of the split beams goes through a polariser, so that
the two photons pumping the $2p$ exciton do not necessarily have the same
polarisation. We show that by rotating one of the polarisers one can switch
on and off the THz laser.

\begin{figure}[tbp]
\vspace{-10pt}
\resizebox{\columnwidth}{!}{\includegraphics{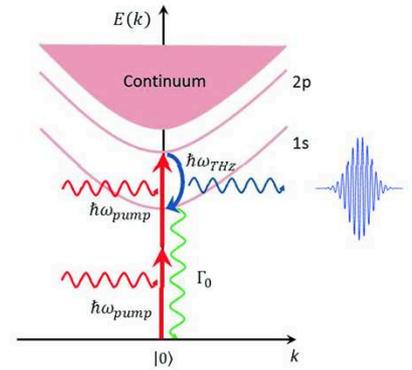}}
\vspace{-40pt}
\caption[Fig 1]{(Color online) Schematic energy-level diagram of two-photon
transitions to $2p$ exciton states in a QW. The ground ($1s$) and excited ($
2p$ dark) discrete (bound) excitonic states and the exciton (unbound)
continuum states are shown. The pumping frequency, $\protect\omega_{pump}$
is half that of the $2p$ exciton state. $\protect\omega_{THz}$ is the center
frequency of the emitted THz pulse; $\Gamma_0$ - ground ($1s$) state exciton
spontaneous emission rate.}
\label{fig:energy_level_pumping_scheme}
\end{figure}

Using crystal symmetry point group theoretical methods \cite{Ivchenko} we
calculate the polarisation dependence of the optical transition matrix
element for two-photon excitonic absorption in GaAs/AlGaAs quantum wells, as
well as of the intra-excitonic $2p$ to $1s$ THz transition radiative decay
rate. This enables us to calculate the polarisation dependence of the
quantum efficiency for THz photon generation and thus identify maximum
efficiency regimes of operation. The optically pumping scheme to a 2p
exciton state by two photons, each of half the energy of the $2p$ exciton
state, is shown in Fig. \ref{fig:energy_level_pumping_scheme}.

\textit{Two-photon $p$-exciton absorption}.- The quasi-2D
(Q2D) exciton wave function at the $\Gamma$ point is given for narrow QWs by \cite{Shimizu}:
\begin{equation}  \label{eq:exciton_wave_function}
\begin{aligned}
& \scalebox{0.85}{\mbox{\ensuremath{\displaystyle \Psi _\lambda  \left( {{\bf
r}_e ,{\bf r}_h } \right) = \frac{{v_0 }}{{\sqrt S }}U_\lambda ^{\alpha
\beta } \left( {\bf \rho } \right)\Phi _c^\alpha  \left( {z_e } \right)\Phi
_v^{\beta *} \left( {z_h } \right)u_{c{\bf k}} \left( {{\bf r}_e }
\right)u_{v{\bf k}}^* \left( {{\bf r}_h } \right)e^{i{\bf k}_{||} .{\bf
R}_{||} }}}}
\end{aligned}
\end{equation}

where $v_0$ is the unit-cell volume, $S$ is the QW area, $\mathbf{R}$ is the
centre-of-mass (c.o.m.) co-ordinate, $\mathbf{r} = \mathbf{r}_e - \mathbf{r}%
_h$ is the relative motion co-ordinate, $\mathbf{r} ({\boldsymbol{\rho}},z)$ and ${\boldsymbol{\rho}}={%
\boldsymbol{\rho}_e}-{\boldsymbol{\rho}_h}$ is the in-plane relative motion
co-ordinate, the $z$ axis is taken normal to the QW layers. $\alpha, \beta$
are subband indices and $\Phi _{c(v)}^\alpha$ is the $\alpha$-subband
envelope function of the conduction (valence) band; $U_\lambda ^{\alpha
\beta } \left( {\mathbf{r}_e - \mathbf{r}_h } \right)$ is the envelope
function of the 2D exciton associated with subbands $\alpha$ of the electron
and $\beta$ of the hole; $\lambda = \left( {n,m} \right)$ is the 2D exciton
quantum number, labelling the discrete excitonic states ($n=1,2,...,$ $%
\left| m \right| < n$); $u_{c\mathbf{k}}, u_{v\mathbf{k}}$ are the periodic
parts of the Bloch wave function for conduction and valence bands,
correspondingly; the exciton c.o.m. wave vector, $\mathbf{k}_{||}\approx 0$, is on the
order of the photon wave vector.
Consider the case of allowed conduction-to-valence band dipole optical
transition at the $\Gamma$ point. In cubic crystals the conservation of
parity upon absorption of two photons requires the final excitonic state to
have the same parity as the valence band, therefore the final exciton is in
a $p$-state. The TPA probability is given by:
\begin{equation}  \label{eq:transition_probability}
W_{TPA} = \frac{{2\pi }}{\hbar }\sum\limits_{if} {\left| {V_{if} } \right|}
^2 S_f \left( E \right)
\end{equation}
where $S_f$ is the final density of states and the momentum, \textbf{p},
matrix element, $V_{if}$, between the initial and final states is given by
\cite{Mahan}:
\begin{equation}  \label{eq:Matrix_element}
\scalebox{0.8}{\mbox{\ensuremath{\displaystyle  V_{fi}  = \frac{{e^2 }}{{m^2
c^2 }}A_1 A_2 \sum\limits_l {\left[ {\frac{{\left\langle f
\right|{\boldsymbol \varepsilon }_{\bf 1} .{\bf p}\left| l \right\rangle
\left\langle l \right|{\boldsymbol \varepsilon }_2 .{\bf p}\left| i
\right\rangle }}{{E_l  - E_i  - \hbar \omega _2 }} + \frac{{\left\langle f
\right|{\boldsymbol \varepsilon }_2 .{\bf p}\left| l \right\rangle
\left\langle l \right|{\boldsymbol \varepsilon }_1 .{\bf p}\left| i
\right\rangle }}{{E_l  - E_i  - \hbar \omega _1 }}} \right]\,\,}}}}
\end{equation}
reflecting the order of absorption of the first photon with polarisation
vector ${\boldsymbol{\varepsilon }}_1$, energy $\hbar \omega_1$ and vector
potential $A_1$ and the second - with polarisation vector ${\boldsymbol{%
\varepsilon }}_2$, energy $\hbar \omega_2$ and vector potential $A_2$. Since
the TPA is a two-step process, one should sum over all intermediate states $%
\left| l \right\rangle$ with energy $E_l$. The first matrix element has been
calculated by Elliott \cite{Elliott} in the 3D case and for the quasi-2D
case here considered reads:
\begin{equation}
\begin{aligned}
& \left\langle l \right|\hat \varepsilon _\alpha .\mathbf{p}\left| i
\right\rangle = \sqrt S \Psi _\lambda ^* \left( 0 \right)\left\langle c
\right|\hat \varepsilon _\alpha .\mathbf{p}\left| v \right\rangle
\\
& = v_0 U_\lambda ^{\alpha \beta } \left( 0 \right)\Phi
_c^\alpha \left( {z_e } \right)\Phi _v^{\beta *} \left( {z_h }
\right)\left\langle c \right|\mathbf{p}\left| v \right\rangle
\end{aligned}
\end{equation}
where $\mathbf{\hat \varepsilon }_\alpha$, $\alpha=1,2$ is the photon
polarisation vector, $\Psi_{\lambda}(0)$ is the relative motion exciton wave
function, given by Eq.(\ref%
{eq:exciton_wave_function}), evaluated at $\mathbf{r=0}$, and the interband matrix element $\left\langle c
\right|\hat \varepsilon _\alpha .\mathbf{p}\left| v \right\rangle$ is
given by:
\begin{equation}
\left\langle c \right|\mathbf{p}\left| v \right\rangle = \frac{1}{{v_0 }}%
\int\limits_{cell} {d^3 ru_c^* \left( \mathbf{r} \right)} \frac{\hbar }{i}%
\nabla u_{v\mathbf{\hat z}} \left( \mathbf{r} \right)
\end{equation}

The second matrix element entering Eq.(\ref{eq:Matrix_element}) is between
hydrogenic-type exciton states and can be written as:
\begin{equation}
\label{eq:second_matrix_element}
\begin{aligned}
& \scalemath{0.85} {\frac{1}{m}\left\langle f \right|\hat \varepsilon _\beta  .{\bf p}\left| l \right\rangle  = \frac{1}{{\bar \mu _\xi  }}\int {d^3 r\,} \Psi _\delta ^{\alpha \beta *} \left( {\bf r} \right)\left( {{\bf \hat \varepsilon }_\beta  .{\bf p}} \right)\Psi _\lambda ^{\alpha \beta } \left( {\bf r} \right)}
\\
& \scalemath{0.85}{= \frac{1}{{\bar \mu _\xi  }}\int {d^3 r\,} U_\delta ^{\alpha \beta *} \left( {\bf \rho } \right)\Phi _c^{\alpha *} \left( {z_e } \right)\Phi _v^\beta  \left( {z_h } \right)\left( {{\bf \hat \varepsilon }_\beta  .{\bf p}} \right)U_\lambda ^{\alpha \beta } \left( {\bf \rho } \right)\Phi _c^\alpha  \left( {z_e } \right)\Phi _v^{\beta *} \left( {z_h } \right)}
\end{aligned}
\end{equation}
where $m$ is the free electron mass and ${\bar{\mu}_{\xi }}$ is the reduced
exciton mass along $\xi $-direction in the QW plane. Introducing a special
notation for the matrix element, summed over the intermediate states,
according to:
\begin{equation}
\label{eq:Integral}
\begin{aligned}
\scalemath{0.85}{
I_\delta  \left( {\alpha ,\beta } \right) =
\int {d^3 r\,} U_\delta ^{\alpha \beta *} \left( {\bf \rho } \right)\Phi _c^{\alpha *} \left( {z_e } \right)\Phi _v^\beta  \left( {z_h } \right)\left( {{\bf \hat \varepsilon }_\beta  .{\bf p}} \right)}
\\
\scalemath{0.85}{
\times \sum\limits_\lambda  {\frac{{U_\lambda ^{\alpha \beta } \left( {\bf \rho } \right)U_\lambda ^{\alpha \beta *} \left( 0 \right)}}{{E_\lambda   + E_G  - \hbar \omega _\alpha  }}} \Phi _c^{\alpha *} \left( {z_e } \right)\Phi _v^\beta  \left( {z_h } \right)\Phi _c^\alpha  \left( {z_e } \right)\Phi _v^{\beta *} \left( {z_h } \right)}
\end{aligned}
\end{equation}
where $E_{G}$ is the direct interband energy gap, the total matrix
element can be written as:
\begin{equation}
\scalebox{0.9}{\mbox{\ensuremath{\displaystyle
V_{fi}  = \sqrt S \frac{{e^2 }}{{m\bar \mu _\xi  c^2 }}A_1 A_2 \left[ {\left\langle c \right|{\bf \hat \varepsilon }_1 .{\bf p}\left| v \right\rangle I_\delta  \left( {1,2} \right) + \left\langle c \right|{\bf \hat \varepsilon }_2 .{\bf p}\left| v \right\rangle I_\delta  \left( {2,1} \right)} \right]}}}
\label{eq:total_matrix_element}
\end{equation}%
Let us define a 2D reduced Coulomb Green's function:
\begin{equation}
G\left( {\boldsymbol{\rho },\boldsymbol{\rho ^{\prime }}}\right)
=\sum\limits_{\lambda }{\frac{{U_{\lambda }^{\alpha \beta }\left( \boldsymbol{%
\rho }\right) U_{\lambda }^{\alpha \beta \ast }\left( {\boldsymbol{\rho }%
^{\prime }}\right) }}{{E_{\lambda }-\Omega }}}
\end{equation}%
where $E_{\lambda }$ is the exciton hydrogenic energy as measured from the
conduction-band edge and $\Omega =-E_{G}+\hbar \omega _{\alpha }<0$. A closed form of the reduced Green's function for
an unscreened exciton in 3D (N-D) has been derived in \cite{Hostler} (\cite{Blinder}) and in the 2D limit of interest is given in \cite{Zimmermann}:
\begin{equation}
\label{eq:Green's_function}
\scalebox{0.9}{\mbox{\ensuremath{\displaystyle
G\left( {{\bf \rho },0} \right) = \frac{1}{{2\pi }}e^{ - {\raise0.7ex\hbox{${2\rho }$} \!\mathord{\left/
 {\vphantom {{2\rho } {\kappa_\alpha  a_B^* }}}\right.\kern-\nulldelimiterspace}
\!\lower0.7ex\hbox{${\kappa_\alpha  a_B^* }$}}} \left[ { - \ln \left( {\frac{{4\rho }}{{\kappa_\alpha  a_B^* }}} \right) - \gamma  + 3 - 4\left( {\frac{\rho }{{\kappa_\alpha  a_B^* }}} \right)} \right]}}}
\end{equation}%
where ${a_{B}^{\ast }}$ is the 3D exciton Bohr radius, $\gamma $ is the
Euler's constant and $\kappa_{\alpha }^{2}=\frac{{E_{B}}}{{E_{G}-\hbar \omega _{\alpha }}}$
with $E_{B}\equiv Ry^{\ast }$ -- the exciton binding energy.\textbf{\ }
Introducing cylindrical co-ordinates and the overlap
integral of the subband envelope wave functions:
\begin{equation}
I_{\alpha \beta }=\left\langle {\Phi _{c}^{\alpha }\left( z\right) }%
\right\vert \left. {\Phi _{v}^{\beta }\left( z\right) }\right\rangle =\int {%
dz}\,\Phi _{c}^{\alpha \ast }\left( z\right) \Phi _{v}^{\beta }\left(
z\right)
\end{equation}%
and momentum matrix element along $z$-direction:
\begin{equation}
P_{\alpha \beta }=\left\langle {\Phi _{c}^{\alpha }\left( z\right) }%
\right\vert p_{z}\left\vert {\Phi _{v}^{\beta }\left( z\right) }%
\right\rangle =\int {dz}\,\Phi _{c}^{\alpha \ast }\left( z\right) \frac{%
\hbar }{i}\frac{\partial }{{\partial z}}\Phi _{v}^{\beta }\left( z\right)
\end{equation}%
the general expression for the sum over intermediate states can be recast
as:
\begin{equation}
\begin{array}{l}
I_{\delta }\left( {\alpha ,\beta }\right) =-i\hbar \frac{{v_{0}^{2}}}{S}%
\left[ {I_{\alpha \beta }\int {d^{2}\rho U_{\delta }^{\alpha \beta \ast
}\left( \rho \right) \left( {\mathbf{\hat{\varepsilon}}_{\beta }.\mathbf{%
\hat{\boldsymbol \rho}}}\right) \frac{{\partial G\left( {\rho ,0}\right) }}{{\partial
\rho }}}}\right.  \\
\left. {+P_{\alpha \beta }\int {d^{2}\rho U_{\delta }^{\alpha \beta \ast
}\left( \rho \right) \left( {\mathbf{\hat{\varepsilon}}_{\beta }.\mathbf{%
\hat{z}}}\right) G\left( {\rho ,0}\right) }}\right]  \\
\end{array}
\label{eq:summation}
\end{equation}%
where $\boldsymbol{\hat{\rho}}=\frac{{\boldsymbol{\rho }}}{{\left\vert {%
\boldsymbol{\rho }}\right\vert }}$ and ${\mathbf{\hat{z}}}$ are unit
vectors.
\begin{figure*}[ht!]
\includegraphics[scale=0.5]{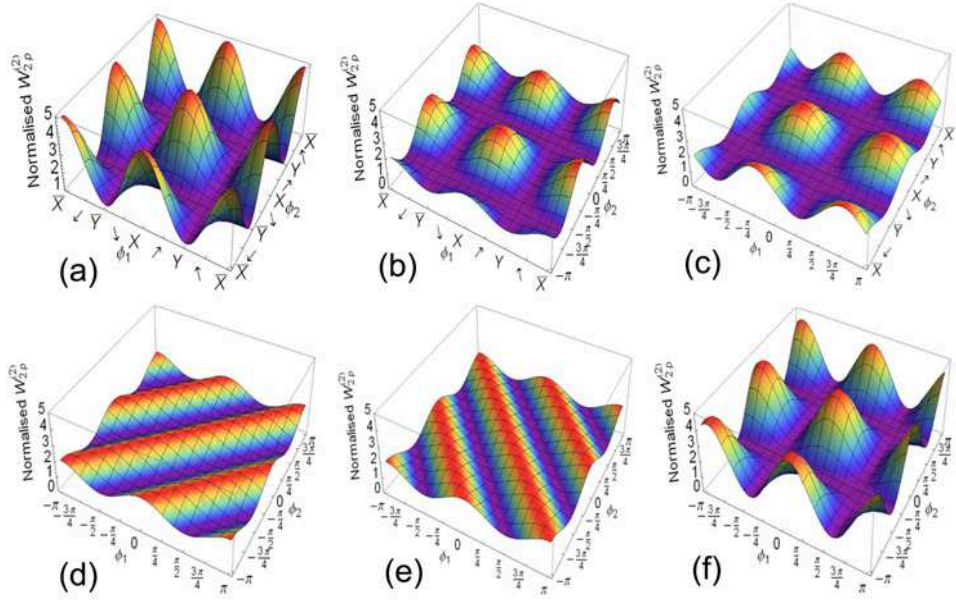}
\vspace{-10pt}
\caption[Fig 2]{(Color online) 3D surface plots of the
normalised excitonic TPA against polar angles of ${\bf \hat \varepsilon }_1$ and ${\bf \hat \varepsilon }_2$
in the QW plane at specific phase shifts $\protect\delta_1,\protect\delta_2$. (a)
co-linearly polarised photons $\protect\delta _{\mathrm{1}} = 0,\pm \protect\pi$;
\,$\protect\delta _2 = 0,\pm \protect\pi$; (b)
$1^{st}$ linear -- $2^{nd}$ $\protect\sigma^{+}$ circularly polarised photon $%
\protect\delta _{\mathrm{1}} = 0;\,\protect\delta _2 = \frac{\protect\pi }{2}
$; (c) $1^{st}$ $\protect\sigma^{+}$ circularly polarised photon -- $2^{nd}$
linearly polarised photon $\protect\delta _{\mathrm{1}} = \frac{\protect\pi
}{2};\,\protect\delta _2 = 0$; (d) $\protect\sigma^{+}$-- $\protect\sigma^{+}$
or $\protect\sigma^{-}$-- $\protect\sigma^{-}$ co-circularly polarised photons
$\protect\delta _{\mathrm{1}} = \pm \frac{\protect\pi }{2};\,\protect\delta %
_2 = \pm \frac{\protect\pi }{2}$; (e) $\protect\sigma^{+}$-- $\protect\sigma%
^{-}$ or $\protect\sigma^{-}$-- $\protect\sigma^{+}$ counter-circularly
polarised photons $\protect\delta _{\mathrm{1}} = \pm \frac{\protect\pi }{2}%
;\,\protect\delta _2 = \mp \frac{\protect\pi }{2}$; (f) co-left-elliptically
polarised photons $\protect\delta _{\mathrm{1}} = - \frac{\protect\pi }{6};\,%
\protect\delta _2 = - \frac{\protect\pi }{8}$.}
\label{fig:TPA_polarisation}
\end{figure*}

For VCSEL configuration and normal incidence
geometry we choose $\mathbf{\hat{\varepsilon}}_{\beta }\bot z$ and
therefore the second term in Eq. (\ref{eq:summation}) vanishes. The derivative of the Green's function can be easily carried out using Eq.(\ref{eq:Green's_function}). We take for the exciton relative motion wave function the 2D hydrogen atom
wave function for bound exciton states \cite{Shinada&Sugano}, \cite%
{Koch&Haug}:
\begin{equation}
\label{eq:2D_exciton_wave_function}
\begin{aligned}
& \scalemath{0.85}{
U_{nm}^{\alpha \beta } \left( {\boldsymbol \rho } \right) = N_{nm}}\left( {\frac{{2\rho }}{{a_B^* \left( {n - \frac{1}{2}} \right)}}} \right)^{\left| m \right|} e^{ - \frac{\rho }{{a_B^* \left( {n - \frac{1}{2}} \right)}}}
\\
& \scalemath{0.85}{
\times L_{n + \left| m \right| - 1}^{2\left| m \right|} \left( {\frac{{2\rho }}{{a_B^* \left( {n - \frac{1}{2}} \right)}}} \right)e^{im\phi }},\,\,\,\,\,\,\,\,n = 1,2,3,...,\left| m \right| < n
\end{aligned}
\end{equation}
where  $N_{nm}= \left[ {\frac{{\left( {n - 1 - \left| m \right|} \right)}}{{\pi a_B^{*2} \left( {n - \frac{1}{2}} \right)^3 \left[ {\left( {n - 1 + \left| m \right|} \right)!} \right]^3 }}} \right]^{1/2}$ and $L_{n}^{\alpha }\left( x\right) $ -- associated Laguerre
polynomials \cite{Abramowitz&Stegun}.

Introducing polar co-ordinates $\left( {\rho
,\varphi }\right)$ we obtain for TPA to $2p$-exciton states
with $n=1, m=\pm 1$:
\begin{equation}
I_{2,1} \left( {\alpha ,\beta } \right) = \frac{{4\hbar I_{\alpha \beta } }%
}{{3\pi i\sqrt {3\pi } \kappa_\alpha a_B^{*3} E_B }}J_{p,2} \left( {\kappa_\alpha }
\right)
\end{equation}
where $E_B = \frac{{\hbar ^2 }}{{2\bar \mu _\xi a_B^{*2} }}$ is the exciton
binding energy and the integral, $J_{p,2} \left( {k_\alpha  } \right)
= - \frac{{9\left( {143 + 36\ln \left( {%
\frac{2}{3}} \right)} \right)}}{{2048}}a_B^{*3} \kappa_\alpha ^3$.

Substituting in Eq. (\ref{eq:total_matrix_element}) the excitonic two-photon absorption matrix element is obtained:
\begin{equation}
V_{fi} = \frac{{v_0^2 }}{\sqrt S }\frac{{e^2 }}{{mc^2 }}A_1 A_2 \frac{{%
4\hbar I_{\alpha \beta } }}{{3\pi i\sqrt {3\pi } }}\frac{1}{{E_B a_B^{*2} }}%
J_{eff,2} \left\langle c \right|\mathbf{p}\left| v \right\rangle
\end{equation}
where we have defined effective matrix element for cubic crystals, using the
invariance of the interband matrix element $M = \left\langle c \right|%
\mathbf{p}\left| v \right\rangle$ under crystal point symmetry group
transformations \cite{Inoue&Toyozawa}, \cite{Mahan}, \cite{Ivchenko}, \cite%
{Heine}:
\begin{equation}  \label{eq:effective_matrix_element}
\begin{aligned}
& \scalemath{0.85}{
J_{eff,2}^2 = \frac{1}{2}\left( {{\bf \hat \varepsilon }_1
\times {\bf \hat \varepsilon }_2 } \right)^2 \left| {J_{p,2} \left( {k_1 }
\right) - J_{p,2} \left( {k_2 } \right)} \right|^2 }
\\
& \scalemath{0.85}{
+ \frac{1}{2}\left[
{1 + \left( {{\bf \hat \varepsilon }_1 .{\bf \hat \varepsilon }_2 }
\right)^2 } \right]\left| {J_{p,2} \left( {k_1 } \right) + J_{p,2} \left(
{k_2 } \right)} \right|^2 }
\\
& \scalemath{0.85}{
= \frac{{C_1^2 }}{2}\left\{ {\left( {{\bf
\hat \varepsilon }_1 \times {\bf \hat \varepsilon }_2 } \right)^2 \left|
{k_1^2 - k_2^2 } \right|^2 + \left[ {1 + \left( {{\bf \hat \varepsilon }_1
.\,{\bf \hat \varepsilon }_2 } \right)^2 } \right]\left| {k_1^2 + k_2^2 }
\right|^2 } \right\}}
\end{aligned}
\end{equation}
where $C_1 = - \frac{{9\left( {143 + 36\ln \left( {\frac{2}{3}} \right)}
\right)}}{{2048}}$.

Our pumping scheme envisages two photons each with half the energy of the $2p
$-exciton state: $\hbar \omega _{1}=\hbar \omega _{2}=\hbar \omega =\frac{{%
E_{2p}}}{2}$ and $k_{1}^{2}=k_{2}^{2}=k^{2}=\frac{{2E_{B}}}{{2E_{G}-E_{2p}}}$%
, therefore the first term in Eq. (\ref{eq:effective_matrix_element})
vanishes and from Eq. (\ref{eq:transition_probability}) we get
for the TPA probability to $2p$-exciton states in [$\mathrm{s^{-1}m^{-2}}$]:
\begin{equation}
\label{eq:two-photon_transition_probability}
\scalebox{0.85}{\mbox{\ensuremath{\displaystyle
W_{2p}^{\left( 2 \right)}  = \frac{{K_{TPA} }}{{27\pi ^3 }}\frac{{C_1^2 }}{2}M^2 I_{\alpha \beta }^2 \left( {\frac{{\hbar ^2 }}{{a_B^{*4} }}} \right)S_{2p}^{c1,hh1} \frac{{16E_B^2 }}{{\left( {2E_G  - E_{2p} } \right)^2 }}\left[ {1 + \left( {{\bf \hat \varepsilon }_1 .{\bf \hat \varepsilon }_2 } \right)^2 } \right]}}}
\end{equation}%
where  $S_{2p}^{c1,hh1}$ is the final $2p$-exciton density of states per unit
area for a heavy-hole exciton (c1-hh1) and the coefficient $K_{TPA}$ for
an infinite quantum well, is given by:
\begin{equation}
K_{TPA}=\frac{{128\pi e^{4}A_{1}^{2}A_{2}^{2}v_{0}^{2}}}{{\hbar \bar{\mu}%
_{\xi }^{2}m^{2}c^{4}SL_{z}^{2}E_{B}^{2}}}
\end{equation}
The photon polarisation vectors ${\mathbf{\hat{\varepsilon}}_{1},%
\mathbf{\hat{\varepsilon}}_{2}}$ with polar angles $\varphi _{1},\varphi _{2}$ and phase shifts $\delta
_{1},\delta _{2}$ correspondingly, lie in the QW plane. 3D plots of the exciton TPA probability are shown in Fig. \ref%
{fig:TPA_polarisation} for different polarisations of the two pumping photons.

We suggest adding an external THz cavity at the VCSEL output that will filter out the linear polarisation of the emitted THz radiation, and will thus constitute our reference frame, fixing the direction of our co-ordinate system x-axis. We shall assume that the generated THz mode is X-polarised. By inspection of Fig. 2 one can see that maximum ($5$-fold) increase of the two-photon absorption rate with respect to $YY$ polarisation is achieved for linearly $XX,X\bar X,\bar XX,\bar X\bar X$ polarised photons
(Fig. \ref{fig:TPA_polarisation}(a)). The two-photon absorption rate can
increase by a factor of $3$ for linearly-circularly or circularly-linearly
polarised photons (Fig \ref{fig:TPA_polarisation}(b,c)); by a factor of $2$ for both circularly polarised (Fig. \ref{fig:TPA_polarisation} (d,e)), by a factor close to $5$ (but
always less than the one for linear polarisation) for elliptically polarised photons (Fig. \ref{fig:TPA_polarisation} (f)).
Our results show that changing polarisation from $YY$ to $XX$, passing through circularly and elliptically polarised pumping, one can vary the lasing threshold by a factor of 5.

\textit{Intra-excitonic $2p \rightarrow 1s$ transition probability}.- We calculate next the polarisation dependence of the $2p \rightarrow 1s$ photon intra-excitonic transition rate, generating THz emission (Fig. \ref
{fig:energy_level_pumping_scheme}). We are interested in the optical
transition matrix element between initial two-fold degenerate state $\Psi_{\delta}$ with $
\delta=(n=2, m=\pm 1)=(2,p)$ and final state $\Psi_{\lambda}$ with $\lambda=(n=1, m=0)=(1,s)$. The matrix element is of the second type Eq.(\ref{eq:second_matrix_element}) and for normal incidence geometry $(\boldsymbol{\hat \varepsilon} \bot \boldsymbol{\hat z})$ and exciton wave functions, given by Eqs. (\ref{eq:exciton_wave_function}),(\ref{eq:2D_exciton_wave_function}), we obtain:
\begin{equation}
\begin{aligned}
& \scalemath{0.75}{
M_{lf} = \frac{{v_0^2 }}{S}\left( {\frac{{\ - 8\,\hbar }%
}{i}} \right)\left( {\frac{m}{{\bar \mu _\xi }}} \right) I_{\alpha \beta }^2 \left( {\frac{1}{{3\sqrt 3 }}} \right)\frac{{%
2\sqrt 2 }}{{3\pi a_B^{*4} }}\Phi \left( \varphi \right) \int {d\rho \,\rho
^2 } e^{ - \frac{{8\rho }}{{3a_B^* }}} L_2^2 \left( {\frac{{4\rho }}{{3a_B^*
}}} \right)}
\end{aligned}
\end{equation}
where we have introduced polar co-ordinates and the angular dependence is
given by: $\Phi \left( \varphi \right) = \cos \varphi \,e^{ \mp i\varphi }$.
The integration over $\rho$ is easily performed, giving: $\frac{{81}}{{512}}%
a_B^{*3}$. Finally, the $2p \rightarrow 1s$ intra-excitonic optical
transition rate for an infinite QW is given by:
\begin{equation}  \label{eq:2p_1s_transition_probability}
W_{2p \to 1s}^{(2)} = \frac{{27}}{{512}}K_{OPA}^{THz} \frac{{\hbar ^2 }}{{%
\pi ^2 a_B^{*2} }}I_{\alpha \beta }^4 S_{1s} \left( E \right)\Phi ^2 \left(
\varphi \right)
\end{equation}
where $S_{1s}(E)$ is the final ($1s$) state density of states and the one-photon THz emission coefficient is given by:
\begin{equation}
K_{OPA}^{THz} = \frac{{16\pi }}{\hbar }\left( {\frac{e}{{\bar \mu_{\xi} c}}}
\right)^2 \frac{{v_0^2 A_{THz}^2 }}{{L_z^4 S}}
\end{equation}
where $A_{THz}$ is the THz photon vector potential, expressed in
terms of the THz emission intensity, $I_{THz}$ as: $A_{THz} = \left( {\frac{{2\pi cI_{THz} }}{{n\omega _{THz}^2 }}} \right)^{1/2}$,
where $n$ is the refractive index and $\omega _{THz} = \frac{{E_{2p} -
E_{1s} }}{\hbar }$.

The polarisation dependence of the THz emission rate can be inferred from the angular dependence: for linear (e.g. along x-axis) polarisation of the emitted THz photon ($\mathbf{\hat \varepsilon } || \,\mathbf{\hat x}$), $\Phi ^2 \left( \varphi \right) = 1$,
for y-linear $\mathbf{\hat \varepsilon } || \,\mathbf{\hat y}$, $\Phi ^2
\left( \varphi \right) = 0$ and therefore there is no THz emission, and for
circularly polarised THz photon, $\Phi ^2 \left( \varphi \right) = \frac{1}{2%
}$, the corresponding THz emission rate is half of the one for x-linear
polarisation.

\textit{Quantum efficiency}.- The quantum efficiency of THz radiation generation can be defined as the
ratio of the THz photon generation rate and the two-photon absorption rate
by a $2p$-exciton and is proportional to the ratio of the squares of the oscillator strengths, $G$ and $g$, of the $2p \rightarrow 1s$ and $|0\rangle \rightarrow |2p\rangle$ transitions \cite{Kavokin}, which can be expressed in terms of the transition
probability \cite{Yariv}:
\begin{equation}
\begin{aligned}
&\scalemath{0.8}{ G^2 \equiv f_{2p \to 1s} = \frac{{6\pi c^3 \varepsilon \bar \mu _\xi S}}{{%
\omega _{THz}^2 n^3 e^2 }}W_{2p \to 1s}^{\left( 2 \right)}}
\\
& \scalemath{0.8}{g^2 \equiv f_{2p} = \frac{{6\pi c^3 \varepsilon \bar \mu _\xi S}}{{\omega
_{2p}^2 n^3 e^2 }}W_{2p}^{\left( 2 \right)}}
\end{aligned}
\end{equation}
where $\omega _{2p} = \frac{{E_{2p} }}{\hbar }$.

Using Eq.(\ref{eq:two-photon_transition_probability}) and Eq.(\ref{eq:2p_1s_transition_probability}),
after some algebra one can obtain for the normalised quantum efficiency:
\begin{equation}
\scalebox{0.9}{\mbox{\ensuremath{\displaystyle
\begin{aligned}
& \scalemath{0.85}{
\eta =\frac{{144m^2 n^2 a_B^{*2} SI_{\alpha \beta }^2 }}{{e^2 \hbar ^2 L_z M^2 \left( {143 + 36\ln \left( {\frac{2}{3}} \right)} \right)^2 }}\frac{{E_{2p}^4 \left( {2E_G  - E_{2p} } \right)^2 }}{{\left( {E_{2p}  - E_{1s} } \right)^3 }}\left( {\frac{{S_{1s}^{c1,hh1} }}{{S_{2p}^{c1,hh1} }}} \right) P\left( {{\bf \hat \varepsilon }_1 ,{\bf \hat \varepsilon
}_2 ,{\bf \hat \varepsilon }} \right)}
\end{aligned}}}}
\end{equation}
where the polarisation dependence is given by:
\begin{equation}
\begin{aligned}
&\scalemath{0.85}{
P\left( {{\bf \hat \varepsilon }_1 ,{\bf \hat \varepsilon
}_2 ,{\bf \hat \varepsilon }} \right)= \frac{{\cos ^2 \left( \varphi \right)}}{{1 +
\left( {\cos \varphi _1 \cos \varphi _2 + \cos \left( {\varphi _1 + \delta
_1 } \right)\cos \left( {\varphi _2 + \delta _2 } \right)} \right)^2 }}}
\end{aligned}
\end{equation}%
and $\varphi $ is the polar angle of the THz photon polarisation vector.
\begin{figure*}[tbp]
\includegraphics[scale=0.28]{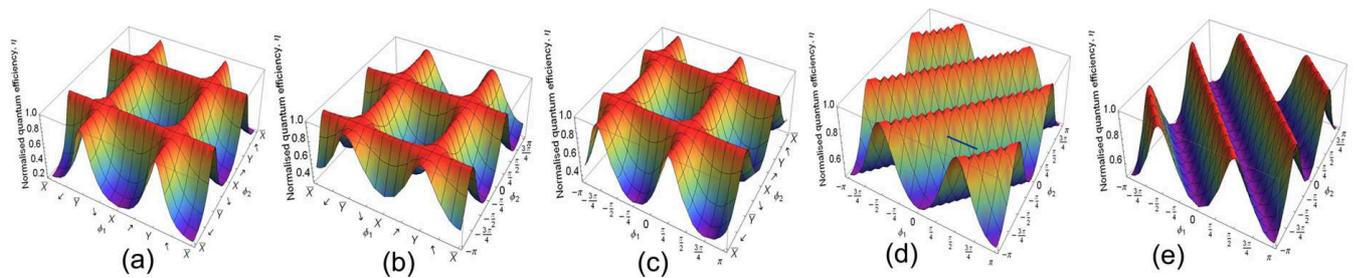}
\caption[Fig 3]{(Color online) 3D surface plots of the
normalised quantum efficiency of THz photon generation against polar angles
of the pumping photons polarisation vectors in the QW plane at different
phase shifts $\protect\delta _{1},\protect\delta _{2}$ at linear, $\protect \phi=0,\pm \pi$ polarisation of the emitted THz radiation (a) co-linearly polarised photons;(b) $1^{st}$ linear-$2^{nd}$ circularly polarised photon; (c) $1^{st}$ circular --$2^{nd}$
linearly polarised; (d) $\protect\sigma ^{+}$--$\protect\sigma ^{+}$ or $\protect\sigma ^{-}$--$\protect\sigma ^{-}$ co-circularly polarised photons; (e) $\protect\sigma ^{+}$--$\protect%
\sigma ^{-}$ or $\protect\sigma ^{-}$--$\protect\sigma ^{+}$
counter-circularly polarised photons.}
\label{fig:quantum_efficiency_polarisation}
\end{figure*}

We shall assume that the $2p$ exciton excited by two-photon absorption has a lifetime, which is long enough that it loses any memory of the polarisation and phase of the excitation, so that it can emit with any polarisation. We shall consider emission with one particular polarisation (either linear or circular) and all possible choices of polarisation of the two pumping photons.

The quantum efficiency polarisation dependence is shown in Fig. \ref{fig:quantum_efficiency_polarisation} for different polarisation
configurations of the two pumping photons at a given (linear) emitted THz photon polarisation. The plots for circular polarisation of the THz radiation look exactly the same but are scaled down by a factor of 2 (not shown), resulting in maximum efficiency $\eta=0.5$. In addition to the results presented in Fig. \ref{fig:quantum_efficiency_polarisation}, we should note that for counter-$X\left( {
\bar{X}}\right)$-linearly polarised pumping photons $\delta_{1}=0\left( \pi \right) ;\delta _{2}=\pi \left( 0\right) $  maximum quantum efficiency $\eta =1$ is achieved for linearly polarised ($\phi =0,\pi $) and $\eta =0.5$ for circularly polarised $(\varphi =\frac{\pi }{4})$ THz emission, unconditionally, for any direction of the linear polarisation of the two pumping photons in the QW plane. Furthermore, if the THz emission is $Y$-linearly polarised, the quantum efficiency $\eta =0$, i.e. no THz radiation should be emitted in this case.
Fig. \ref{fig:quantum_efficiency_polarisation} shows that the maximum quantum efficiency $\eta = 1$ could be achieved within certain regions in the plane for linearly polarised THz emission for all combinations of linear and circular polarisations of the two pumping photons. Note that in both (circular and linear THz emission polarisation) cases, maximum quantum efficiency is achieved along $YY$ lines for co-linearly polarised photons,
for $Y$-polarised first (second) photon in the case of linear-circular (circular-linear) polarisation, or along diagonal lines for co- and counter- circular-circular polarisation of the pumping photons. We emphasise, however, that although maximum quantum efficiency could be achieved both by counter- and co-linearly polarised photons, the quantum efficiency in the former case is constant and does not depend on the direction of the polarisation vectors in the QW plane, while maximum quantum efficiency in the latter case is obtained solely for specific directions of the polarisation vectors in the plane ($YY$).

In order to verify these predictions experimentally, one can envisage
pumping of a QW structure with two laser beams having the same
frequency (equal to a half of the $2p$-exciton resonance frequency) but
different polarisation. These two beams may be generated by the same laser
but should propagate through different polarisers before focussing on the sample.
In addition we suggest including a delay line between the two parts of the pumping beam, which would provide the
phase difference of $\pi$ between them to obtain counter-linearly polarised beams for which unconditional maximum efficiency is predicted. The intensity and polarisation of the THz light emitted by the structure could
be measured as a function of intensities and polarisations of the two pumping
beams. As reference experiments one can measure the intensity of THz
emission with one of the pump beams switched off. Analysing the results of such
experiments one should bear in mind that the two photons used to generate a $2p$%
-exciton may originate from the same beam as well as from different beams.
Comparing the spectra obtained with both beams switched on with those
obtained with only the first or only the second beam switched on, one can
extract the signal generated by absorption of the two photons coming from
different beams and thus having different polarisations.

\textit{Conclusions}.- We have developed a theory of the two-photon
absorption to $p$-exciton states in QWs and calculated the polarisation
dependence of two-photon transition probability, using crystal symmetry
point group methods. We show that the two-photon transition rate is strongly
dependent on the polarisation of both photons and our model predicts variation of the lasing threshold by a factor of $5$
by switching from $YY$ to  co-linearly $XX$-polarised
pumping. We calculated the polarisation dependence of the intra-excitonic THz
emission and the quantum efficiency for THz photon generation. Maximum quantum efficiency
is predicted for counter-linearly polarised pumping photons and linearly polarised THz emission.
Conditions for achieving maximum quantum efficiency for different
polarisations of the pumping photons are identified, thereby opening routes for
polarisation control of the THz VCSEL and a range of new applications
entailed from it.

We thank Prof. E. L. Ivchenko for valuable discussions. AK acknowledges
financial support from the EPSRC Established Career Fellowship grant.






\newpage

\end{document}